\documentclass[traditabstract,a4,10pt]{aa}

\usepackage{graphicx}
\usepackage{natbib,amsmath,amssymb,xcolor}

\newcommand{\fedge}{f_{\rm edges}}
\newcommand{\res}{{\rm err}}

\newcommand{\atan}{\,{\rm atan \;} }

\newcommand{\elik}{{\mathbf K}}

\newcommand{\ain}{{a_{\rm in}}}
\newcommand{\aout}{{a_{\rm out}}}

\newcommand{\gzpacz}{{g_Z^{\rm Pacz.}}}

\newcommand{\km}{k_-}
\newcommand{\kp}{k_+}
\newcommand{\kpm}{k_\pm}

\newcommand{\minnerdisc}{M_{\rm inner \, disc}}

\begin{document}

\title{Self-gravity in thin discs and edge effects: an extension of Paczynski's approximation}

\titlerunning{Self-gravity in thin discs and edge effect}

\author{Audrey Trova\inst{^{1,2}}, Jean-Marc Hur\'e\inst{1,2}, \and Franck Hersant\inst{^{1,2}}}

\offprints{audrey.trova@obs.u-bordeaux1.fr}

\institute{Univ. Bordeaux, LAB, UMR 5804, F-33270, Floirac, France
\and
CNRS, LAB, UMR 5804, F-33270, Floirac, France\\
\email{audrey.trova@obs.u-bordeaux1.fr}\\
\email{jean-marc.hure@obs.u-bordeaux1.fr}\\
\email{franck.hersant@obs.u-bordeaux1.fr}}

\date{Received ??? / Accepted ???}
 
\abstract{As hydrostatic equilibrium of gaseous discs is partly governed by the gravity field, we have estimated the component caused by a vertically homogeneous disc, with a special attention for the outer regions where self-gravity classically appears. The accuracy of the integral formula is better than $1 \%$, whatever the disc thickness, radial extension and radial density profile. At order zero, the field is even algebraic for thin discs and writes $- 4 \pi G \Sigma(R) \times \fedge (R)$ at disc surface, thereby correcting Paczynski's formula by a multiplying factor $\fedge \gtrsim \frac{1}{2}$, which depends on the relative distance to the edges and the local disc thickness. For very centrally condensed discs however, this local contribution can be surpassed by action of mass stored in the inner regions, possibly resulting in $\fedge \gg 1$. A criterion setting the limit between these two regimes is derived. These result are robust in the sense that the details of vertical stratification are not critical. We briefly discuss how hydrostatic equilibrium is impacted. In particular, the disc flaring should not reverse in the self-gravitating region, which contradicts what is usually obtained from Paczynski's formula. This suggests that i) these outer regions are probably not fully shadowed by the inner ones (important when illuminated by a central star), and ii) the flared shape of discs does not firmly prove the absence or weakness of self-gravity.
}


\maketitle

\section{Introduction}

Self-gravity in a disc orbiting a central star is important if it is massive enough or/and very extended \citep{ab84,mohanty13}. It can influence not only its rotation curve and trigger instabilities (including large scale patterns and fragmentation), but also the accumulation of matter around the mid-plane. Actually, the perpendicular component of the field governs | together with the temperature | the disc thickness, internal pressure, density, etc. In a seminal paper, \cite{pacz78} has studied the impact of vertical self-gravity on the hydrostatic equilibrium of a Keplerian disc. He assumed that the gradients of the potential are mainly vertical, making the Poisson equation fully integrable by analytical means, like for plane parallel sheets \citep{gl65}. With such an hypothesis, the field is linear with the surface density, which enables to treat discs as concentric rings, independent to each other. Paczynski's approximation was subsequently considered by many authors in the context of star formation and accretion in Active Galactic Nuclei through ``one zone'' and bi-dimensional computations \citep{pacz78bis,sc81,sw82,cgw82,sb87,lx90,c92,cr92,fs92,hureclp94,lxj94,lxj95,mu96,mu97,mnu97,pg98,bhd98,hure98,hure00,dsb00,hrz01,vb03,md05,kd08}. Depending on the viscosity prescription \citep{ss73,pringle81}, models predict the rise of the density along with the disc flattening and even pinching, which conditions are suited to the birth of global gravitational instabilities \citep{armitage2010astrophysics}.

Paczynski's approximation is expected to break down in certain zones where gradients of density and thickness are noticeable. These are typically disc edges, and possibly internal gaps and transition regions where matter experiences intrinsic changes \citep[e.g., variation of opacity, or mean molecular weight; see][]{pdl97,hure00,npmk00}. Like any open boundary, edges are important components of discs both for geometrical and dynamical reasons (significant area, boundary layer, reservoir of angular momentum, escape of radiation, contraction/expansion into the ambient medium, etc.). In many models and grid-based hydrodynamical simulations, edges are often located outside the computational box, and are supposed not to be influential \citep{ss73,cd90,dubrulle92}, which requires specific boundary conditions \citep[e.g.][]{pbm11}. Unfortunately, there is no easy way to determine with precision how the gravitational field varies at edges, except from a three-dimensional numerical approach. In this paper, we reconsider Paczynski's formula for vertical self-gravity. Our main objective is to extend its domain of validity, in particular at the outer edge (i.e., large radii) where the field gradually falls and self-gravity can be important. It is actually important to improve the understanding of those regions where most observations come from. As a matter of fact, this study also warns that Paczynski's formula is never valid in the common context of astrophysical discs, which questions models released earlier (see references above). We reduce the technical difficulties by making a few major assumptions, namely: i) axial and mid-plane symmetries, ii) small aspect ratio, and iii) constant density profile in the direction perpendicular to the mid-plane. This framework is therefore well suited to ``vertically averaged, geometrically thin'' discs. Thus, we derive an extended formula, discuss its limits and accuracy. In particular, by considering a few realistic density profiles, we conclude that only the local surface density is critical, not the vertical stratification. We then briefly discuss how the extended formula should impact hydrostatic equilibrium. Clearly, it would be worth exploring its consequences by performing a detailed disc model, which could be the aim of another article.  

The paper is organised as follows. In Section \ref{sec:localvsnonlocal}, we recall the expression for the vertical field in an infinite plane-parallel medium. As a matter of fact, the monopole approximation is fairly not relevant for cases of astrophysical interest. We show through a few concrete examples that Paczynski's approximation fails: i) close to disc edges, and ii) for centrally condensed discs. The estimate of the vertical field begins at Section \ref{sec:eeffect} from the exact the integral expression valid for vertically homogeneous discs. We first simplify the axially symmetric kernel (which involves complete elliptic integrals) with a $3$-term fit (accuracy of $1\%$ typically). We then use this fit to derive a formula appropriate for geometrically thin discs and valid for a wide range of radial density profiles and disc's shape. Next, we derive a criterion stating the condition of non-local contributions (i.e., the case of centrally condensed discs). We show that this result is compatible with Paczynski's formula in the limit of infinite, plane-parallel sheets. In Section \ref{sec:ozm}, we discuss edge effects. We consider power-law disc models, give a zero order correction to Paczynski's formula, check the assumption of vertical homogeneity, and stress how self-gravity should significantly change hydrostatic equilibrium and disc flaring. The last section is devoted to concluding remarks.

\section{Local vs. non-local character of self-gravity}
\label{sec:localvsnonlocal}

\subsection{Plane-parallel sheets and Paczynski's approximation}

\begin{figure}
\includegraphics[width=8.9cm]{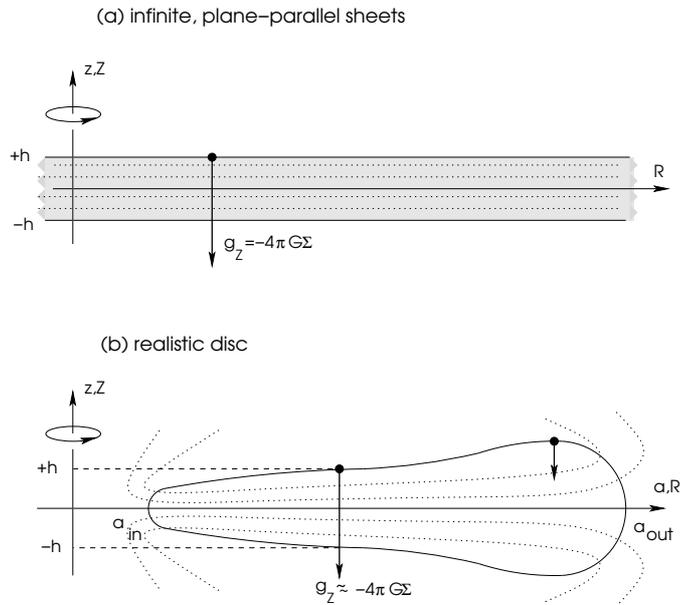}
\caption{In the infinite, plane-parallel sheet model (a), the magnitude of vertical self-gravity $g_Z$ is fully determined from the Poisson equation (or Gauss theorem) and is linear with the cumulative surface density. In real systems (b), this is only correct in magnitude (Paczynski's approximation): important density gradients or large concentrations of mass in the inner disc produce  significant deviations to this law.}
\label{fig:img.eps}
\end{figure}

In cylindrical coordinates $(R,\theta,Z)$, the Poisson equation linking the gravitational potential $\psi$ to the mass density $\rho$ writes
\begin{equation}
\frac{1}{R}\frac{\partial }{\partial R} \left(R \frac{\partial \psi}{\partial R}  \right) + \frac{1}{R^2} \frac{\partial^2 \psi}{\partial \theta^2} + \frac{\partial^2 \psi}{\partial Z^2} =4\pi G \rho.
\label{eq:poisson}
\end{equation}
When the mass density is constant in any $(R,\theta)$-plane, the radial and azimuthal gradients of the potential are strictly zero everywhere. The medium is then made of plane-parallel and homogeneous sheets \citep{gl65}. The configuration is depicted in Fig. \ref{fig:img.eps}a. Equation (\ref{eq:poisson}) can then be integrated in the vertical direction from the mid-plane $Z=0$, leading to
\begin{equation}
\left.\frac{\partial \psi}{\partial Z}\right|_Z - \left. \frac{\partial \psi}{\partial Z} \right|_0 = 4 \pi G \Sigma(Z),
\label{eq:vertgrad}
\end{equation}
where
\begin{equation}
\Sigma(Z) = \int_{0}^{Z}{\rho(z) dz}
\end{equation}
is the cumulative surface density, i.e. the mass of gas per unit area from the midplane to $Z$. If the medium is symmetric with respect to the mid-plane, then $\left. \partial \psi/\partial Z \right|_0=0$, and the vertical component of gravity $g_z=-\partial_Z \psi$ writes
\begin{equation}
g_Z(Z) = - 4 \pi G \Sigma(Z),
\label{eq:isa}
\end{equation}
which result can also be established from Gauss theorem \citep{armitage2010astrophysics}.  As long as the following conditions \citep[hereafter ``Paczynski's approximation'';][]{pacz78}
\begin{equation}
\begin{cases}
\frac{1}{R}\frac{\partial }{\partial R} \left(R \frac{\partial \psi}{\partial R}  \right) \ll  \frac{\partial^2 \psi}{\partial Z^2},\\\\
\frac{1}{R^2} \frac{\partial^2 \psi}{\partial \theta^2} \ll \frac{\partial^2 \psi}{\partial Z^2},
\label{eq:negl_grads}
\end{cases}
\end{equation}
are satisfied, Eq.(\ref{eq:vertgrad}) roughly holds, but the field is only approximate, i.e.
\begin{equation}
g_Z(Z)  \approx - 4 \pi G \Sigma(Z) \equiv \gzpacz.
\label{eq:gzpaszynski}
\end{equation}

\begin{figure}
\includegraphics[width=8.9cm, bb=18 38 709 522, clip=]{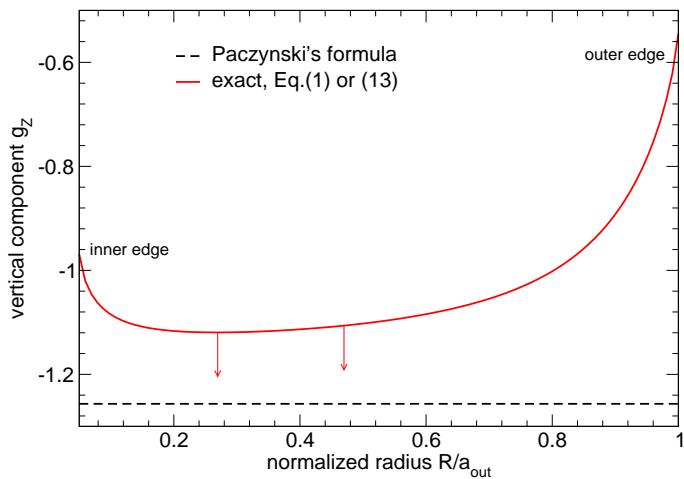}
\caption{Vertical component $g_Z$ of self-gravity of a homogeneous disc computed from Eq.(\ref{eq:gz1}) compared to Paczynski's formula versus the radius, for $Z=h$ (i.e., at the top of the disc). The disc parameters are: $\ain/\aout=0.05$, mass density $\rho=1$, and semi-thickness $h=0.1 \aout$ (model 1 in Tab \ref{tab:3models}). When the radial extent of the disc increases, $g_z$ tends to Paczynski's formula, as indicated by the arrows but edge effects are still present.}
\label{fig:check.eps}
\end{figure}

\begin{figure}
\includegraphics[width=8.9cm, bb=28 38 709 529, clip=]{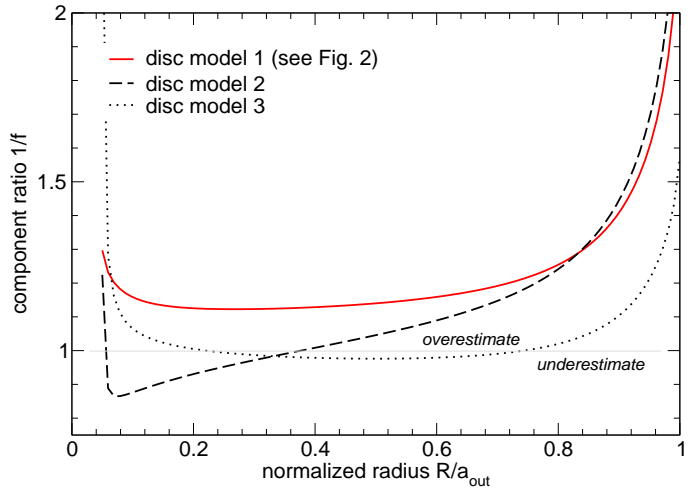}
\caption{Component ratio $1/f$ defined by Eq.(\ref{eq:f}) for the three disc models defined in Tab. \ref{tab:3models} including the case shown in Fig. \ref{fig:check.eps} (the disc parameters are the same). When $f < 1$, Paczynski's formula overestimates the field.}
\label{fig:check_err.eps}
\end{figure}

In fact, this expression is not very accurate in discs, which can generally not be considered as plane-parallel and homogeneous sheets. Discs are finite in size and are inhomogeneous. A simple proof is shown in Fig. \ref{fig:check.eps} where we compare the ``exact'' vertical field\footnote{In the paper throughout, the reference fields are computed from Eq.(\ref{eq:gzgeneral}) through the splitting method \citep{hure05}, which is equivalent to solving Poisson equation.} at the top of a thin, finite size disc with constant density and constant semi-thickness $h$ (disc model 1), with the value given by Eq.(\ref{eq:gzpaszynski}). The deviations are the largest at edges. Paczynski's formula overestimates the field by a factor $2$ typically close to the outer edge: there is no matter beyond the outer edge, which weakens the field. Obviously, the deviation is reduced at intermediate radii (i.e. far from edges) when the disc axis ratio $\ain/\aout$ decreases, but edge effects are always present. These concern a few local thicknesses in radius. Figure \ref{fig:check_err.eps} shows the ratio
\begin{equation}
\frac{\gzpacz}{g_Z} \equiv  \frac{1}{f}
\label{eq:f}
\end{equation}
for the three disc models listed in Tab. \ref{tab:3models}, including the fully homogeneous case considered before. Thus, $f < 1$ means that Paczynski's formula overestimates the field. We see that Eq.(\ref{eq:gzpaszynski}) always overestimates the field at large radii.

\subsection{Centrally condensed discs}

\begin{figure}
\includegraphics[width=8.9cm,bb=0 150 520 650, clip=]{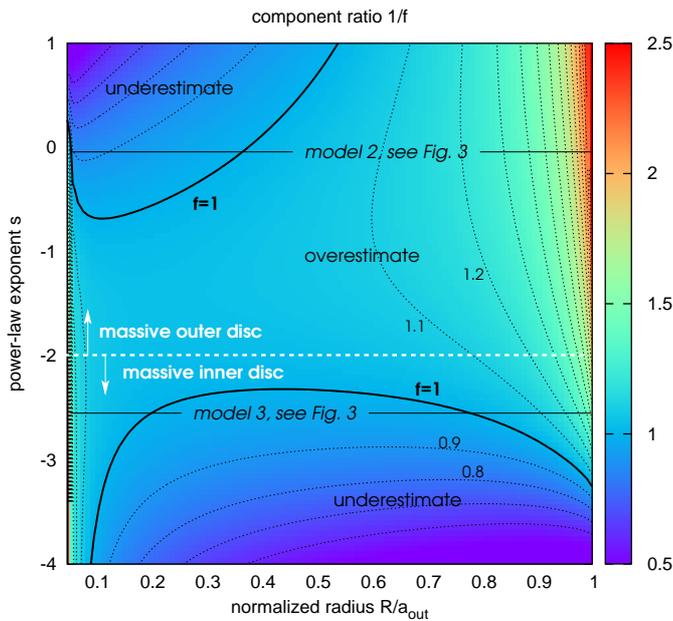}
\caption{Component ratio $1/f$ defined by Eq.(\ref{eq:f}) for the same parameters as for model 3 defined in Tab. (\ref{tab:3models}) but different power-law exponents $s$ in the range $[-4,1]$ (contour lines every $0.1$).}
\label{fig:err.eps}
\end{figure}

 For centrally condensed discs (like model 3), a non-local effect appears. Actually, if the mass contained in the inner disc is large enough, then the outer disc feels the field of a central condensation, and so
\begin{equation}
g_Z  \approx - \frac{G\minnerdisc}{R^2} \times \frac{Z}{R}.
\label{eq:gzcentrallycondensed}
\end{equation}
where $\minnerdisc$ is close to the total disc mass. Since this non-local field is not limited in magnitude, it can surpass the local contribution, and Paczynski's value as well, resulting in $f>1$ in Eq.(\ref{eq:f}). To check this point, we have considered model 3 but for different power-law density profiles, namely
\begin{equation}
\rho(a) \propto a^s,
\label{eq:plaw}
\end{equation}
where the exponent $s$ is a constant. Figure \ref{fig:err.eps} displays $1/f$ versus $R$ and for a continuum of the power-law exponents $s \in [1,-4]$, which should concern most cases of astrophysical interest. Here, the disc aspect ratio
\begin{equation}
\epsilon=\frac{h(a)}{a}
 \end{equation}
is set to $0.1$, and the axis ratio is the same as for Fig. \ref{fig:check.eps}. We clearly see that, for $s \lesssim -2$, the gravity of the inner disc can not be ignored. This depends on $\epsilon$ and $\ain/\aout$.

\begin{table}
\centering
\begin{tabular}{cccl}
model & $h$ &  $s$ & comment\\ \hline 
1 & $0.1 \aout$ & $0$ & homogeneous and flat disc\\
2 & $0.1 a$ & $0$ & model 1 but constant aspect ratio\\
3 & $0.1 a$ & $-2.5$ & model 2 but {\bf radially} inhomogeneous\\
  \hline
\end{tabular}
\caption{Three disc models used in this study: $h$ is the semi-thickness and $s$ is the exponent of the power-law profile for the mass density, i.e., $\rho(a) \propto a^s$.}
\label{tab:3models}
\end{table}

We conclude that the presence of edges makes the use of Paczynski's approximation incorrect by a factor $0.5-2$ typically. Two contributions can play a role: the local disc, and the inner regions if these contain a large fraction of the disc mass. It would be erroneous to consider this range is not wide enough to require a more detailed analysis. As quoted above, the hydrostatic equilibrium of a disc, through the equation
\begin{flalign}
\nonumber
 g_Z & = \frac{1}{\rho}\partial_Z P\\
     & \approx - \frac{1}{h} \times \left. \frac{P}{\rho}\right|_{Z=0}
\label{eq:gzcentrallycondensed}
\end{flalign}
where $P$ is the pressure, is very sensitive to gravity. We see that an error by a factor $2$ in the field results in a factor $\sqrt{2}$ on the gas temperature.

\section{Vertical gravity of vertically homogeneous discs}
\label{sec:eeffect}

\begin{figure}
\begin{center}
\includegraphics[width=8.5cm,bb=35 35 709 522,clip=]{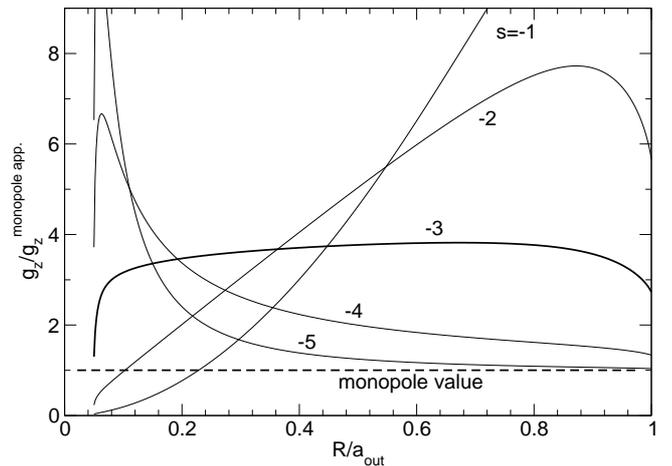}
\end{center}
\caption{Ratio of the vertical component of the disc gravity to the monopole value versus the radius for model 3 but different power law exponents $s$ of the density. Same edges as for Fig. \ref{fig:check.eps}.}
\label{fig:cc_vs_kep.eps}
\end{figure}

The component we wish to calculate and compare to Eq.(\ref{eq:gzpaszynski}) is determined from the integral: 
\begin{equation}
g_Z(\vec{r})=-G \iiint_{\rm disc}{\frac{\rho(\vec{r}') (\vec{r}-\vec{r}').\vec{e}_Z a da d\theta' dz}{|\vec{r}-\vec{r}'|^3} },
\label{eq:gzgeneral}
\end{equation}
where $\vec{r}'(a,\theta',z)$ refers to matter and $\vec{e}_Z$ is the unit vector. This integral is too complex to be estimated without additional hypothesis. A series expansion of the kernel which leads to a component of the form
\begin{equation}
- G m(R)\frac{Z}{R^3} + \dots,
\label{eq:monop}
\end{equation}
where $m(R)$ is the mass of gas from the inner edge of the disc to radius $R$, could be envisaged but it remains inaccurate in reproducing the field, especially for exponent power-law $s \gtrsim -4$ \citep{mu97,fa12}. This is illustrated in Fig. \ref{fig:cc_vs_kep.eps} which displays the exact component $g_z$ at the top of the disc divided by Eq.(\ref{eq:monop}) at the same altitude versus the radius, for flared, power law discs with exponents $ \in [-5,-1]$. We see that this ``monopole-type'' approximation is valid only for very condensed discs with $s \lesssim -5$. The above integral approach is therefore fully justified.

Under axial and mid-plane symmetries, and for $\partial_z\rho=0$, the integrations over $z$ and $\theta'$ in Eq.(\ref{eq:gzgeneral}) can be performed analytically. We have \citep{durand64,hure05}:
\begin{equation}
\label{eq:gz1}
g_Z=-2G \int_a{\rho \sqrt{\frac{a}{R}} [\kp\elik(\kp)-\km\elik(\km)] da},
\end{equation}
where
\begin{equation}
\kpm=\frac{2\sqrt{aR}}{\sqrt{(a+R)^2+(Z\mp h)^2}} \in [0,1]
\end{equation}
is the modulus of the complete elliptic integral of the first kind $\elik(k)$, and $\pm h$ is the altitude of the top/bottom surface of the disc. Both $\rho$ and $h$ can depend on the radius $a$. We note that the integrand in Eq.(\ref{eq:gz1}) is zero at the mid-plane (i.e., for $Z=0$), whatever $a$ and $R$ since $k_+=k_-$. Off the mid-plane, the function $k \elik(k)$ exhibits a peak at $a=R$; this peak rises as $Z \rightarrow \pm h$, and is even infinite when $Z=\pm h$, which corresponds to $k=1$ (the elliptic integral is logarithmically singular; see the Appendix). This peak lends weight to the mass density distributed around $R$, and is therefore responsible for the {\it local character} of self-gravity. This function has also broad wings, which are important if $\rho$ happens to be large at radii $a \ll R$ or $a \gg R$. This is a {\it non-local effect}.

\begin{figure}
\begin{center}
\includegraphics[width=8.5cm,bb=48 47 709 530,clip=]{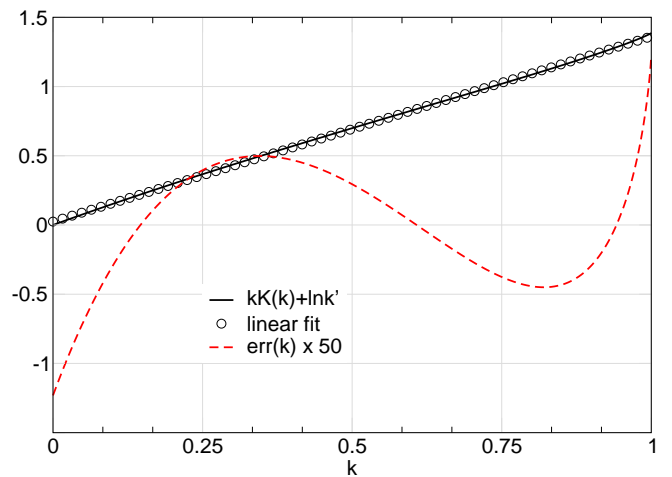}
\end{center}
\caption{Function $k\elik(k)+\ln k'$ versus $k$ and its linear fit by Eq.(\ref{eq:3termfit}). The absolute error is also shown (dotted line).}
\label{fig:fitkK.eps}
\end{figure}

\begin{figure}
\includegraphics[width=9.1cm,bb=0 150 540 650, clip=]{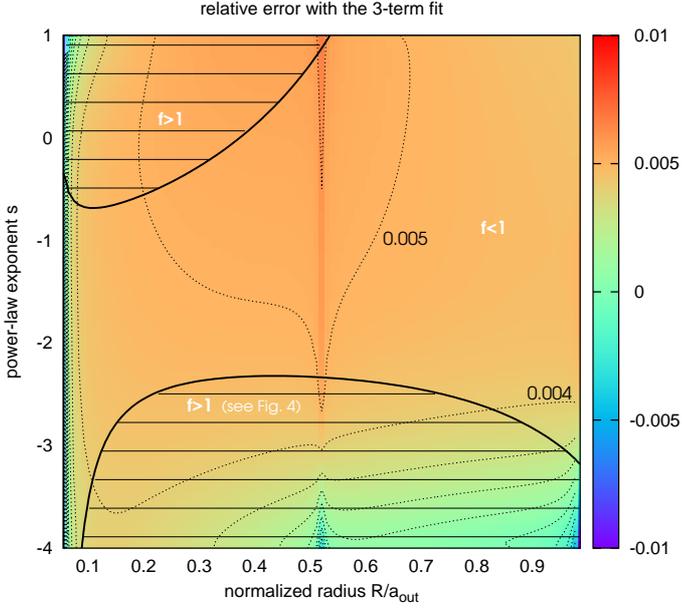}
\caption{Relative deviation between the vertical component $g_Z$ computed from the three-term fit of the function $k \elik(k)$ and from the exact kernel $k \elik(k)$. The conditions are the same as for Fig. \ref{fig:err.eps} (contour dotted lines every $0.001$).}
\label{fig:err_3fit.eps}
\end{figure}

\subsection{A three-term fit of the kernel}

To estimate Eq.(\ref{eq:gz1}) by analytical means, we must replace the special function by a series. Expanding $\elik(k)$ gives at the lowest order \citep{gradryz07}:
\begin{equation}
\begin{cases}
\frac{2}{\pi}\elik(k) \approx 1 + \frac{k^2}{4} \quad \text{for} \quad k \rightarrow 0,\\\\
\elik(k) \approx \ln \frac{4}{k'} \quad \text{for} \quad k \rightarrow 1,
\end{cases}
\end{equation}
where $k'=\sqrt{1-k^2}$ is the complementary modulus. These two asymptotic forms do not connect together well at intermediate modulus however. The five-term and nine-term fits given in \cite{as70} are not very practical and honestly too accurate for the present purpose. We have noticed that the function $k\elik(k)+\ln k'$, when plotted versus $k$, is very close to linear. This is shown in Fig. \ref{fig:fitkK.eps}. We have then considered the following three-term fit:
\begin{equation}
\label{eq:3termfit}
k\elik(k) = C_0+C_1k - \ln k' + \res(k),
\end{equation}
where $\res(k)$ is the error. The result of a least-square regression gives $C_0=0.0246$ and $C_1=1.3371$. Figure \ref{fig:fitkK.eps} shows the associated errors. The area under the curve is remarkably conserved as we have
\begin{flalign}
\nonumber
\int_0^1{\res(k)dk} &= 1 - \left(C_0+\frac{1}{2}C_1-\ln 2+1\right)\\ & \approx 2 \times 10^{-6}.
\end{flalign}
 This approximation is surely not perfect\footnote{Other pairs $(C_0,C_1)$ can probably be used here. For instance, with $C_0=0$ and $C_1=\ln 4$, the three-term fit becomes exact both at $k=0$ and $k=1$, and the integral of the error is strictly zero, but the least-square fit is less satisfactory.} but it minimises the subsequent analytical treatment and provides good results since $|\res(k)| \lesssim 0.015$ in the whole domain $k \in [0,1]$. Using Eq.(\ref{eq:3termfit}), the integrand in Eq.(\ref{eq:gz1}) is
\begin{equation}
\kp\elik(\kp)-\km\elik(\km) \approx -C_1 (\kp-\km)- \ln \frac{\kp'}{\km'},
\label{eq:fittokK}
\end{equation}
with an accuracy of $1 \%$, or less. This is proved in Fig. \ref{fig:err_3fit.eps} which gives the relative error made on the vertical component using this three-term fit instead of the exact kernel. The conditions are the same as for Fig. \ref{fig:err.eps}. Note that coefficient $C_0$ does not play a role in the following. 
 
\subsection{Application to geometrically thin discs}

At this level, we can estimate the vertical field for any type of vertically homogeneous medium by inserting the three term fit in Eq.(\ref{eq:gz1}). We find
\begin{equation}
\label{eq:gz1with3termfit}
g_Z=+2G \int_a{\rho \sqrt{\frac{a}{R}} \left[C_1 (\kp-\km) + \ln \frac{\kp'}{\km'}\right] da},
\end{equation}
which only avoids manipulating special functions (see the Appendix for the numerical evaluation). We can simplify further the formula by assuming
\begin{equation}
(Z \mp h)^2 \ll (a+R)^2,
\end{equation}
which is suited to discs such that $\epsilon^2 \ll 1$ and close neighborhood. Equation (\ref{eq:gz1with3termfit}) can then be written in the form
\begin{equation}
g_Z  \approx -G \int_a{\rho(a) w_1(a) da} - G\int_a{\rho(a) w_2(a) da},
\label{eq:gzw1w2}
\end{equation}
where
\begin{equation}
\begin{cases}
w_1(a)= -\sqrt{\frac{a}{R}}\ln \frac{(a-R)^2+(Z-h)^2}{(a-R)^2+(Z+h)^2} > 0,\\\\
w_2(a)= -4Zh\sqrt{\frac{a}{R}}\frac{1-m C_1}{(a+R)^2},
\end{cases}
\end{equation}
and
\begin{equation}
m=\frac{2\sqrt{aR}}{(a+R)} \in [0,1].
\end{equation}

\begin{figure}
\begin{center}
\includegraphics[width=8.9cm,bb=47 38 709 522,clip=]{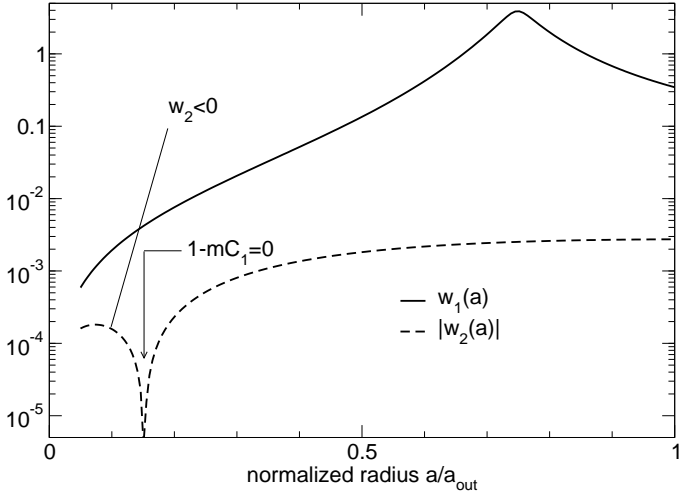}
\end{center}
\caption{The two weighting functions $w_1$ and $w_2$ plotted versus the normalised radius $a/\aout$ for model 2 (see Tab. \ref{tab:3models}), $\epsilon=0.1$, $Z=\frac{3}{4}h$ and $R/\aout=0.75$.}
\label{fig:w1w2.eps}
\end{figure}

Figure \ref{fig:w1w2.eps} displays  $w_1$ and $w_2$ versus $a/\aout$ in a typical case. We see that $w_1$ dominates in absolute close to the outer edge. This is true for any altitude between the mid-plane and the surface $Z=\pm h$. For $a \ll R$ however, $w_1$ and $w_2$ are close in magnitude, meaning that $w_2$ can probably not be neglected for centrally condensed discs.

\subsection{Local vs. non-local contributions: a criterion}

As stated above, the local contribution of self-gravity is due to the peak of the function $w_1(a)$ at $a=R$, while the possible role of the inner disc is due to the left wing of $w_1$. We can derive a limit between these two regimes by equating the local and non-local contributions. To avoid the possible singularity at $a=R$, we integrate $w_1(a)$ over one semi-thickness leftward and rightward to $R$, and then divide by $2h$. To compute this radial average, we assume that $\rho(a)$ and $h(a)$ do not change significantly in the integration range $a \in [R-h(a),R+h(a)]$. After calculation, we find
\begin{flalign}
\frac{1}{2h}\int_{-h}^h{w_1(a)da}&=\ln 4+4 \atan \frac{1}{2}\\
& \equiv A \approx 3.241,
\nonumber
\end{flalign}
at $Z=h$. For $a \ll R$, still assuming small radial gradients for $\rho(a)$ and $h(a)$, we have roughly
\begin{equation}
\frac{1}{2h}\int_{-h}^h{w_1(a)da} \approx w_2(a).
\end{equation}
Provided $\rho(a)$ decreases monotonically outward, the local contribution exceeds the non-local (i.e., inner disc) contribution by a factor $\alpha$ if
\begin{equation}
A \rho(R) \gtrsim \alpha \rho(a) w_1(a),
\end{equation}
which leads to the criterion
\begin{equation}
\frac{A}{4\alpha} \frac{\rho(R)}{\rho(a)} \ge \sqrt{\frac{a}{R}}\frac{h(R)h(a)}{R^2}.
\label{ineq:crit}
\end{equation}
A safe criterion is then obtained with $\alpha \sim 10$ (see Sect. \ref{sec:ozm} for power-law discs).

\subsection{Recovering Paczynski's formula ?}
\label{subsec:pacz}

To verify that  Eq.(\ref{eq:gzw1w2}) is compatible with Paczynski's formula, we must include appropriate assumptions, namely $\rho(a)=cst$ and $h(a)=cst$. Next, we omit $w_2$, which is much less than $w_1$ in magnitude. We then have
\begin{equation}
g_Z \approx -G \rho(R) \int_a{w_1(a) da}.
\end{equation}
Setting $\sqrt{\frac{a}{R}}  \approx 1 + \frac{a-R}{2R}$, the order zero term is
\begin{flalign}
\label{eq:gzedge}
\int{w_1(a) da} & \approx \int{\ln \frac{(a-R)^2+(Z-h)^2}{(a-R)^2+(Z+h)^2} da}\\
\nonumber
&= (a-R) \ln \frac{(a-R)^2+(Z-h)^2}{(a-R)^2+(Z+h)^2}\\
& + 2\left[(Z-h)\atan{\frac{a-R}{Z-h}} - (Z+h)\atan{\frac{a-R}{Z+h}}\right].
\nonumber
\end{flalign}
We now take $R- nh$ and $R+nh$ as integral bounds, where $n$ is a positive number. The contribution of matter located leftward to $R$ is
\begin{flalign}
\nonumber
\int_{a=R}^{a=R-nh}{w_1(a) da}& \approx 4Z \frac{n}{1+n^2} - 2(h-Z) \atan \frac{nh}{h-Z}\\
&+ 2(h+Z) \atan \frac{nh}{h+Z}.
\end{flalign}
Assuming $n \gg 1$, this gives
\begin{equation}
\int_{a=R}^{a=R-nh}{w_1(a) da} \approx 2 \pi Z.
 \end{equation}
Since the integrand in Eq.(\ref{eq:gzedge}) is odd in $a-R$, the right side contribution is the same. We finally find
 \begin{equation}
g_Z \approx -G \rho(R) \times 2 \pi Z \times 2,
\end{equation}
which is equivalent to Eq.(\ref{eq:gzpaszynski}) for a vertically homogeneous slab. This demonstration clearly explains why the presence of an edge suppresses one wing in $w_1(a)$, which diminishes the vertical field by a factor $\sim 2$ typically.

\subsection{Is vertical structure critical ?}
\label{subsec:zstrcuture}

\begin{figure}
\includegraphics[width=9.cm,bb=20 38 710 530, clip=]{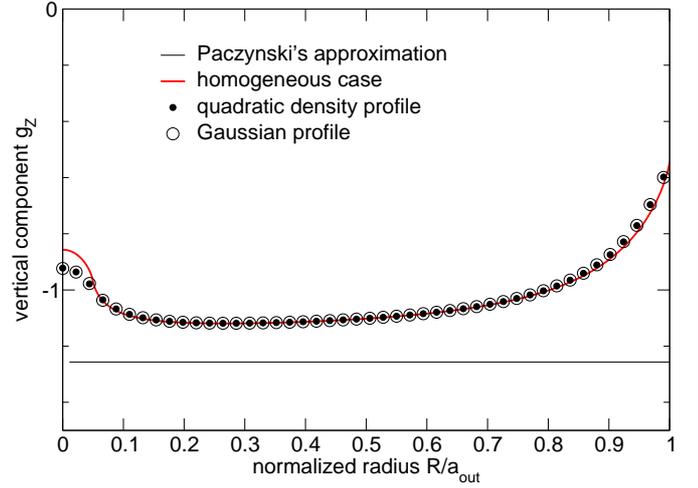}
\caption{Same legend and same conditions as for Fig. \ref{fig:check_err.eps} (disc model 1), but for the Gaussian and quadratic profiles. Deviations with respect to the vertically homogeneous case are globally of the order of the percent or less.}
\label{fig:verticalstrat0.eps}
\end{figure}

\begin{figure}
\includegraphics[width=9.cm,bb=20 38 710 530, clip=]{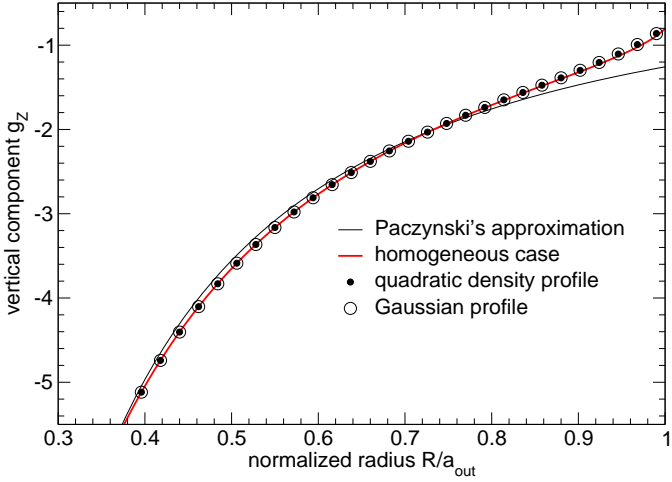}
\caption{Same as for Fig. \ref{fig:verticalstrat0.eps} but for model 3.}
\label{fig:verticalstrat.eps}
\end{figure}

The results presented so far are valid for vertically homogeneous systems only. In order to check the sensitivity to this hypothesis which may seem too restrictive for applications, we have considered two density profiles:
\begin{equation}
\rho(z)=\rho_0 \times
\begin{cases}
\frac{2}{\sqrt{\pi} {\rm erf}({\sqrt{2})}}e^{-2z^2/h^2},\\\\
\frac{3}{2}\left(1-\frac{z^2}{h^2}\right).
\end{cases}
\end{equation}
The first one is a classical Gaussian distribution, typical of vertically isothermal discs, and the second one its quadratic version which, in contrast, has a vanishing density at the surface. Both have the same cumulative surface density at $z=h$, i.e. $\rho_0h$, so the comparison of the vertical fields with the homogeneous case is straightforward. The results are displayed in Fig. \ref{fig:verticalstrat0.eps} for model 1, and in Fig. \ref{fig:verticalstrat.eps} for model 3 (with a zoom at the outer edge). The deviations caused by a non-uniform vertical stratification are in fact of the order of the percent and smaller. Note that, in the Gaussian case, matter located at $z\ge h$ ($\sim 5\%$ of the total mass per unit area) has almost no influence. We can conclude that all formulae derived in this paper assuming $\partial_z \rho=0$ at any radius are relatively robust.

\section{Measuring edge effects}
\label{sec:ozm}

\subsection{Zero-order correction to Paczynski's formula}

\begin{figure}
\includegraphics[width=9.1cm,bb=0 150 530 650, clip=]{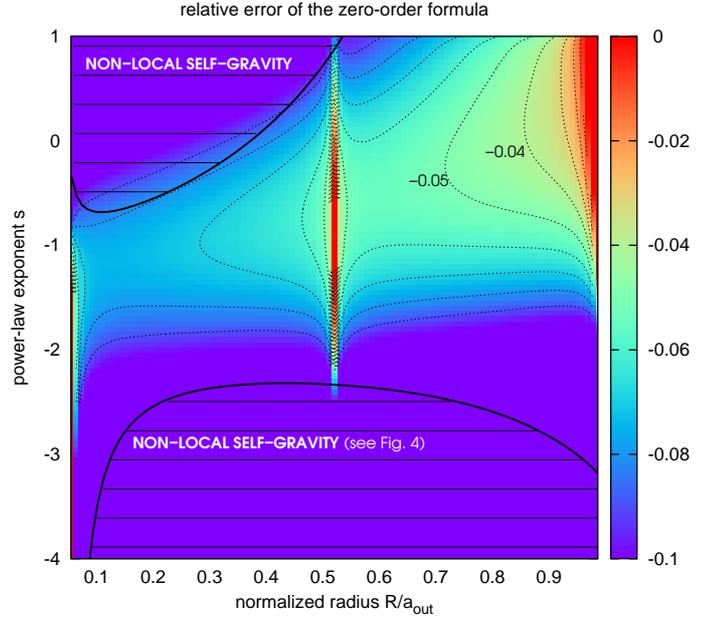}
\caption{Relative error on the vertical field when computed from Eq.(\ref{eq:extendedpacz}) (i.e. the zero order extension of Paczynski's formula), for the same parameters as for model 3 defined in Tab. (\ref{tab:3models}) but different power-law exponents $s$ in the range $[-4,1]$ (contour lines every $0.01$).}
\label{fig:err_order0.eps}
\end{figure}

It is not easy to deduce an algebraic formula for $g_Z$ at the top of the disc, even when the local contribution is dominant. However, we can get the order zero by considering the development above. If we define the variable
\begin{equation}
\varpi=\frac{|a-R|}{2h(R)},
\end{equation}
which measures the distance to the actual radius $R$ in units of local thickness $2h(R)$, then Eq.(\ref{eq:gzedge}) gives:
\begin{equation}
\label{eq:gzedge_orderzero}
\int{w_1(a) da} \approx 4h(R) \left[ \varpi \ln \frac{\varpi}{\sqrt{1+\varpi^2}}-  \atan \varpi \right],
\end{equation}
where we have set $Z=h(R)$. The vertical field at the surface of the disc is then given, for $R \in [\ain,\aout]$, by the relation 
\begin{equation}
\label{eq:extendedpacz}
g_Z \approx \gzpacz \times \fedge,
\end{equation}
where
\begin{equation}
\fedge = f_{\rm in} + f_{\rm out},
\end{equation}
\begin{equation}
\begin{cases}
\pi f_{\rm in} &=  \atan \varpi_{\rm in} - \varpi_{\rm in} \ln \frac{\varpi_{\rm in}}{\sqrt{1+ \varpi_{\rm in}^2}}, \\\\
\pi f_{\rm out} &= \atan \varpi_{\rm out} - \varpi_{\rm out} \ln \frac{\varpi_{\rm out}}{\sqrt{1+ \varpi_{\rm out}^2}} ,
\end{cases}
\end{equation}
and $\varpi_{\rm in}$ is for $\ain$, and $\varpi_{\rm out}$ is for $\aout$. Clearly, $\fedge$ is a function of the radius $R$. Figure \ref{fig:err_order0.eps} shows the accuracy of this zero-order approximation (i.e. the relative error made on $\fedge$), in the same conditions as Figure \ref{fig:err.eps}. We see that the error does not exceed $10 \% \sim \epsilon$ in the domain where the local contribution dominates (i.e. for $s \gtrsim s_0$). If the mass stored in the inner discs does not play a role, we have $\pi f_{\rm in} \approx \frac{\pi}{2}$, and so
\begin{equation}
\fedge \approx \frac{1}{2} + \atan \varpi_{\rm out} - \varpi_{\rm out} \ln \frac{\varpi_{\rm out}}{\sqrt{1+ \varpi_{\rm out}^2}} .
\end{equation}

\begin{figure}
\begin{center}
\includegraphics[width=8.9cm,bb=28 38 709 522,clip=]{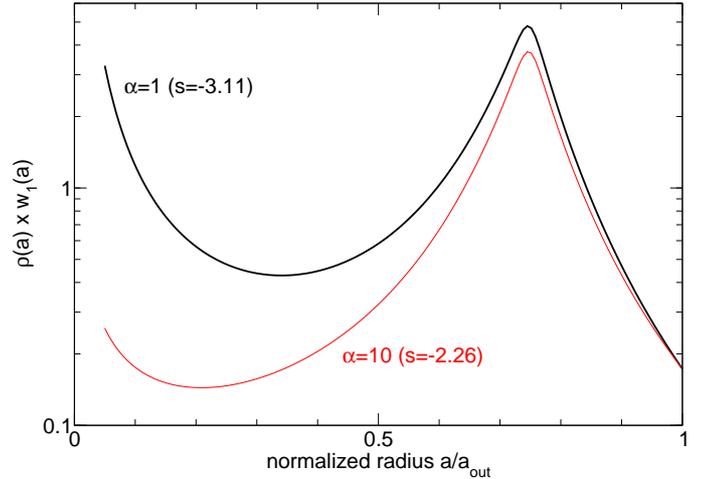}
\end{center}
\caption{The function $\rho(a) \times w_1(a)$ plotted versus the normalised radius $a/\aout$ in the same condition as for Fig. \ref{fig:w1w2.eps}, but for two different exponents: $s=-2.26$ (which corresponds to $\alpha=10$) and for $s=-3.11$ (which corresponds to $\alpha=1$).}
\label{fig:criterion311.eps}
\end{figure}

\subsection{The criterion for power-law discs}

\textcolor{red}{For} power-law discs where $\rho(a)$ is defined by Eq.(\ref{eq:plaw}) and constant aspect ratio $\epsilon$, Ineq.(\ref{ineq:crit}) becomes
\begin{equation}
s \gtrsim 
 - \frac{3}{2} + \frac{\ln \frac{4 \alpha \epsilon^2}{\pi}}{\ln \frac{R}{\ain}}.\label{eq:critpld}
\end{equation}

In the conditions of Fig. \ref{fig:w1w2.eps}, the criterion $s \gtrsim -2.26$ for $\alpha=10$, i.e., the local contribution is larger than the inner edge contribution by a factor $10$. This is $s \gtrsim -3.11$ for $\alpha=1$. We see in Fig. \ref{fig:criterion311.eps}, where we have plotted $w_1(a) \times \rho(a)$ versus $a$, that the criterion is reliable. To render the criterion robust and independent on $R$, we can set $R = \aout$ and $\alpha=10$. Then, the criterion reads
\begin{equation}
s \ge -1.5 + \frac{2 \log \epsilon+1.1}{\log \aout/\ain}.
\end{equation}

\subsection{Are self-gravitating discs still flaring ?}

\begin{figure}
\begin{center}
\includegraphics[width=8.9cm,bb=28 46 709 522,clip=]{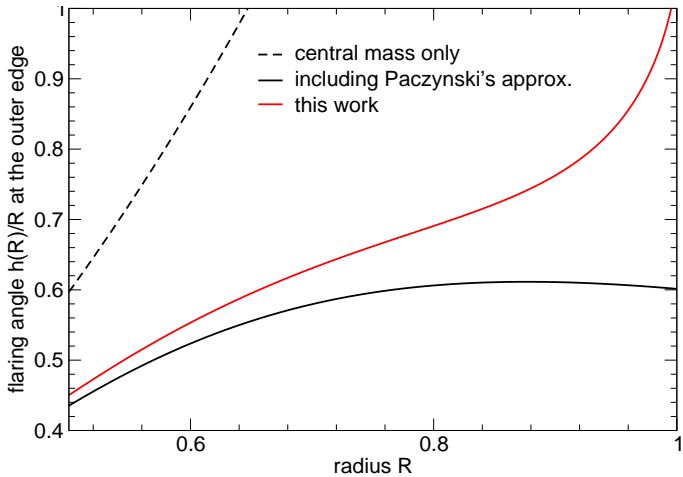}
\end{center}
\caption{Flaring angle $\epsilon(R)$ close the disc outer edge for various models of vertical field including a central star ten times more massive than the disc. The disc parameters other than the semi-thickness $h$ are the same as for Fig. \ref{fig:check.eps} (see text).}
\label{fig:hoverr.eps}
\end{figure}

How the disc shape is eventually modified in the self-gravitating region ? It is not possible to give a firm answer to this question without considering a detailed disc model including balance and transport equations. As quoted in the introduction, models usually show that the thickness is reduced when Paczynski's formula is applied. The fact that $g_Z/\Sigma \propto \fedge$ decreases close to the edge implies a thickening of the disc. Figure \ref{fig:hoverr.eps} compares the local flaring angle $\epsilon(R)$ in a disc close to the outer edge when using Paczynski's formula and when using our new expression. In this simple example, the temperature $\sim P/\rho$ is kept constant and we just compare $h/R$ resulting from Eq.(\ref{eq:gzcentrallycondensed}) with two different formulae for $g_Z$. We also include the contribution of a central star such that the disc mass is $10\%$ the central mass. We recognize the disc pinching, typical of Paczysnki's formula. This property is not found by considering the edge effect. On the contrary, the disc is expected to exhibit a certain flaring, less than in a Keplerian case, but a flaring. Obvisouly, a detailed disc model is required to take into account any decrease of $\Sigma$ and $T$ outward.

\section{Conclusion}

In this article, we have revisited Paczynski's formula for vertical self-gravity in order to extend its domain of validity. The primary aim is to furnish a reliable analytical tool to better investigate the structure of outer parts of discs, through ``one-zone'' models typically. We have demonstrated that the classical approximation is erroneous by a factor $0.5-2$ typically in the conditions of common use, which can have a noticeable influence on the disc equilibrium. Even, for very centrally condensed discs, it can underestimate gravity much more. Our formula works for a wide family of radial profiles and shape, and its accuracy is better than $1 \%$ in any case. Depending on the disc thickness (low or large), different formulae can be implemented in models. Interestingly enough, the details of the vertical density structure is not important, only the surface density at the top of the disc is critical. For geometrically thin discs however, the vertical field is, at zero order at least, fully algebraic. If necessary, high-orders can be considered.

There are a few interesting consequences to the decrease of vertical self-gravity close to the disc edge. As noticed, the disc could not be pinched as previously concluded (see references in the introduction), but could still flare. Obviously, this claim must be adjusted with the necessary transition region from the disc to the ambient/external medium, through a decrease of the surface density and  which temperature. Nevertheless, the edge effect could render the disc thicker than thought, and subsequently less dense. This adresses two interesting issues. Can the outer disc still be illuminated by a central star if present \cite[e.g.][]{dd04} ? Can the decrease of the density postpone the triggering of global instabilities and affect Toomre's criterion ? There could be a range of disc masses, high enough to be subject to vertical self-gravity, but low enough to prevent from global instabilities.

\begin{acknowledgements}
It is a pleasure to thank the referee, W.J. Duschl, for valuable comments and suggestions, in particular about vertical stratification.
\end{acknowledgements}

\bibliographystyle{aa}

\onecolumn
\appendix

\section{Estimation of $\int_a{\sqrt{\frac{a}{R}}\rho(a) k\elik(k) da}$ through integral splitting}

We can estimate numerically Eq. (\ref{eq:gz1}) with accuracy by using integral splitting \citep{ph04,hure05}. In the present case, we extract the log. singularity from the function $\elik$, leaving its regular part, denoted $\elik^*$. Then, we decompose the integrand as follows
\begin{flalign}
\sqrt{\frac{a}{R}}\rho(a) k\elik(k) &=\sqrt{\frac{a}{R}}\rho(a) k\left[ \elik^*(k)-\ln k'\right] \\
& = \sqrt{\frac{a}{R}}\rho(a) k \elik^*(k)-\sqrt{\frac{a}{R}} k \rho(a) \ln k' \\
& = \sqrt{\frac{a}{R}}\rho(a) k \elik^*(k)-\left[\sqrt{\frac{a}{R}} k \rho(a) - \rho(R) \right]\ln k' -  \rho(R) \ln k' \\
& = \sqrt{\frac{a}{R}}\rho(a) k \elik^*(k)-\left[\sqrt{\frac{a}{R}}k\rho(a) - \rho(R) \right]\ln k'  -  \rho(R) \left\{\ln k' - \ln \sqrt{\frac{(a-R)^2+[Z-h(R)]^2}{(a+R)^2+[Z-h(R)]^2}} \right\} \\
\nonumber
& \qquad\qquad -  \frac{1}{2}\rho(R) \ln \frac{(a-R)^2+[Z-h(R)]^2}{(a+R)^2+[Z-h(R)]^2} 
\end{flalign}

We see that the first three terms are always finite, even when $k=1$. The last term is possibly singular, but it contains terms of the form $cst \times \ln (x^2 +cst)$ which are integrable by analytical means. This method also works to estimate Eqs.(\ref{eq:gz1with3termfit}) and (\ref{eq:gzw1w2}).

\end{document}